\documentclass[10pt]{iopart}
\usepackage{epsfig}
\usepackage{graphicx}
\usepackage{dcolumn}
\usepackage{iopams}
\usepackage{color}

\expandafter\let\csname equation*\endcsname\relax 
\expandafter\let\csname endequation*\endcsname\relax 

\usepackage{amsmath}
\usepackage[english]{babel}
\usepackage[utf8]{inputenc}
\usepackage{hyperref}
\usepackage{bm}

\begin{document}

\title{Spectral Gaps of Spin-orbit Coupled Particles in Deformed Traps}

\author{O V Marchukov, A G Volosniev, D V Fedorov, A S Jensen and N T Zinner}
\address{Department of Physics and Astronomy, Aarhus University, 
DK-8000 Aarhus C, Denmark}

\begin{abstract}
We consider a spin-orbit coupled system of particles in an external trap that 
is represented by a deformed harmonic oscillator potential. The spin-orbit 
interaction is a Rashba interaction that does not commute with the trapping 
potential and requires a full numerical treatment in order to obtain the 
spectrum. The effect of a Zeeman term is also considered. 
Our results demonstrate that variable spectral gaps occur as a function of 
strength of the Rashba interaction and deformation of the harmonic trapping 
potential. The single-particle density of states and the critical strength for 
superfluidity vary tremendously with the interaction parameter. The strong 
variations with Rashba coupling and deformation implies that the few- and 
many-body physics of spin-orbit coupled
systems can be manipulated by variation of these parameters.
\end{abstract}
\date{\today}
\maketitle

\section{Introduction}
The last decade is associated with breakthroughs in ultracold atomic
and state-of-the-art optical lattice experiments, which provided 
not only extremely important data for understanding of the fundamentals 
of quantum mechanics, but also introduce a sophisticated set of tools for 
investigation of exceptionally pure and tunable quantum systems \cite{ketterle2008,bloch2008,esslinger2010,cirac2012}.
Recently, this toolbox has been expanded to also include the possibility
of applying controllable gauge fields to both bosonic \cite{lin2009a,lin2009b,lin2011,aidelsburger2011,zhang2012} 
and
fermionic systems \cite{wang2012,cheuk2012} (see references~\cite{dalibard2011} or \cite{zhai2012} for short reviews). 
Spin-orbit coupling is a prime example of a non-abelian
potential that plays an important role throughout physics. Examples
are the spin-orbit splittings in atomic and nuclear spectra, and the 
distinct effect it imposes on the band structure of solid-state 
systems that has lead to great recent advances in the exploration 
of materials with robust metallic surface states, the so-called 
topological insulators \cite{hasan2010,qi2011}.

In many experiment with ultracold atoms, the external trapping 
potentials, while present, are often quite shallow and can 
be ignored for many purposes or one can include them using 
a local density approximation. However, some recent experiments
have demonstrated that tight trapping is not only possible but
also a very exciting possibility as this brings the physics one
can study close to what is known from atomic or nuclear structure.
Some recent theoretical papers have shown that the trapping potential
can play an important role in spin-orbit coupled systems 
\cite{ghosh2011,sinha2011,liu2012,hu2012}. The cited works 
consider the case where the trapping potential is isotropic 
in all three dimensions \cite{ghosh2011} or isotopic in the 
plane where the spin-orbit coupling acts \cite{sinha2011,liu2012,hu2012,larson2013}.
In the current presentation we extend this discussion by considering 
also the case where the trapping potential is deformed. This can 
be done by adjusting the optical or magnetic trap geometry \cite{bloch2008}.

Our findings show that deformation of the external potential 
has a decisive effect on the spectral density and can lead
to the opening and (near) closing of the energy gaps in the 
single-particle spectrum. In the strongly deformed 
effectively one-dimensional limit the level spacing 
is completely determined by the shallow trap harmonic
frequency at low energy. However, for small deformation 
one can find regimes where the spectral gaps will 
almost close and produce a (quasi)-continuum. This 
implies that deformation can be used as a control parameter
for the level density. In the case of a many-body
system with non-abelian gauge potentials, 
the spectral density plays a  
decisive role in trapped systems when exploring many-body phenomena
such as exotic pairing and crossover \cite{vya2011,yu2011,gong2011,iskin2011,vya2012}, superfluidity and 
condensation \cite{stanescu2008,wang2010,ho2011,struck2012,zhou2012}, 
ferromagnetism \cite{baur2012} or 
quantum Hall states \cite{goldman2007,goldman2009,anderson2010,cooper2011,grass2011,goldman2012,hauke2012,beugeling2012}.
Our results represent an initial step in exploring these 
interesting questions for spin-orbit coupled systems in 
deformed traps.

\section{Formalism}
We consider the single-particle problem in a deformed three-dimensional (3D) 
harmonic trap including a Rashba \cite{rashba1960} spin-orbit interaction and an
applied magnetic field. Here we use the Rashba form of the 
spin-orbit interaction but the formalism is completely general and 
applies to all types of spin-orbit coupling. This includes the case where the Rashba-type and 
Dresselhaus-type \cite{dresselhaus1955} contributions are equal (amounting to 
a term, $\sigma_x p_y$, in the notation introduced below which is used in many current cold atomic gas experiments studying spin-orbit effects).
In experiments with cold atomic gases, the spin-orbit coupling is produced by laser beams and 
the spin-orbit coupling strength can be controlled by the laser intensity \cite{dalibard2011}. This gives a large degree
of tunability that is outside the current scope of analogous systems in condensed-matter physics \cite{juzeliunas2010}.

The Hamiltonian for a particle with mass $m$ experiencing a Rashba and Zeeman term is
\begin{equation}\label{hamil}
H = \frac{{\hat p_x}^2}{2m} + \frac{1}{2}m \omega_{x}^2 {\hat x}^2 + \frac{{\hat p_y}^2}{2m} + \frac{1}{2}m \omega_{y}^2 {\hat y}^2 + \frac{{\hat p_z}^2}{2m} + \frac{1}{2}m \omega_{z}^2 {\hat z}^2 + \alpha_R (\hat \sigma_x \hat p_y - \hat \sigma_y \hat p_x) - \bm{\mu}\cdot\mathbf{B},
\end{equation}
where $\hat \sigma_x$ are $\hat \sigma_y$ are $2 \times 2$ Pauli matrices and $\alpha_R$ is the strength of the Rashba spin-orbit coupling. We consider a particle with two internal degrees of freedom, which we label as spin up $\mid \uparrow \rangle$ and spin down $ \mid \downarrow \rangle$. We choose the effective magnetic field as $\bm{B} =(0,0,B)$ and we can write the 
Zeeman term as $-\mu B \hat \sigma_z$ with $\hat \sigma_z$ the remaining Pauli matrix and $\mu$ is the effective magnetic moment.
For the harmonic trapping potentials we assume different trapping frequencies in all directions, $\omega_x$, $\omega_y$ and $\omega_z$. However, 
since we take the Rashba term to act only in the $xy$-plane the $z$-direction effectively decouples from the problem. Note that 
in the case of $B=0$, the Hamiltonian is symmetric under time-reversal and for a single fermion Kramers theorem dictates
a two-fold degeneracy.

The Hamiltonian can be written in the matrix form
\begin{equation}
\begin{pmatrix}
\hat H_{0x} + \hat H_{0y} - \mu B  & \alpha_R (\hat p_y + i \hat p_x) \\
\alpha_R (\hat p_y - i \hat p_x) & \hat H_{0x} + \hat H_{0y} + \mu B
\end{pmatrix}
\begin{pmatrix}
\psi_\uparrow \\
\psi_\downarrow
\end{pmatrix}
= E 
\begin{pmatrix}
\psi_\uparrow \\
\psi_\downarrow
\end{pmatrix},
\end{equation}
where we have introduced the short-hand $\hat H_{0x}=\tfrac{{\hat p_x}^2}{2m} + \tfrac{1}{2}m \omega_{x}^2 {\hat x}^2$ and similarly for $\hat H_{0y}$. This 
constitutes an effective two-dimensional (2D) problem. 
The oscillator energy in z-direction can be considered as a parameter that
can be included in the total energy, $E$. 
The natural basis in which to expand our wave-functions are eigenfunctions of 
the 2D harmonic oscillator. Thus, for a 
\begin{align}
&\psi_{\uparrow} = \sum_{n_x, n_y} a_{n_x, n_y} \mid n_x, n_y,\uparrow \rangle \\&
\psi_{\downarrow} = \sum_{n_x, n_y} b_{n_x, n_y} \mid n_x, n_y,\downarrow  \rangle,
\end{align}
where $\mid n_x, n_y \rangle$ are vectors in the Hilbert space of the two-dimensional harmonic oscillator solutions.

Introducing the standard ladder operators $\hat a_{x,y}$ and $\hat a_{x,y}^{\dagger}$ that obey
\begin{align}
\hat x = \sqrt{\frac{\hbar}{2m\omega_x}}(\hat a_{x}^{\dagger} + \hat a_{x}), \hspace{1 cm} \hat p_x = i\sqrt{\frac{m\hbar\omega_x}{2}}(\hat a_{x}^{\dagger} - \hat a_{x}) \nonumber, \\
\hat y = \sqrt{\frac{\hbar}{2m\omega_y}}(\hat a_{y}^{\dagger} + \hat a_{y}), \hspace{1 cm} \hat p_y = i\sqrt{\frac{m\hbar\omega_y}{2}}(\hat a_{y}^{\dagger} - \hat a_{y}),
\end{align}
the Hamiltonian matrix now reads
\begin{equation}
\begin{pmatrix}
\gamma(\hat{N}_x + \frac{1}{2}) + (\hat{N}_y + \frac{1}{2}) - \frac{\mu B}{\hbar\omega_y}  & 
\beta(i(\hat a_{y}^{\dagger} - a_{y}) - \sqrt{\gamma}(\hat a_{x}^{\dagger} - a_{x})) \\
\beta(i(\hat a_{y}^{\dagger} - a_{y}) + \sqrt{\gamma}(\hat a_{x}^{\dagger} - a_{x})) & \gamma(\hat{N}_x + \frac{1}{2}) + (\hat{N}_x + \frac{1}{2}) + \frac{\mu B}{\hbar\omega_y}
\end{pmatrix}
\begin{pmatrix}
\psi_\uparrow \\
\psi_\downarrow
\end{pmatrix}
= \varepsilon
\begin{pmatrix}
\psi_\uparrow \\
\psi_\downarrow
\end{pmatrix},
\end{equation}
where we have introduced the notation $\hat{N}_x=\hat a_{x}^{\dagger}\hat a_{x}$ and $\hat{N}_y=\hat a_{y}^{\dagger}\hat a_{y}$.
We have here defined the ratio, $\gamma$, of the harmonic oscillator frequencies as $\gamma=\tfrac{\omega_x}{\omega_y}$ and the 
dimensionless Rashba coupling $\beta = \alpha_R \sqrt{\frac{m}{2\hbar \omega_y}}$. Note that the latter implies that we measure the velocity $\alpha_R$ in units of the harmonic oscillator velocity $\sqrt{\frac{\hbar \omega_y}{m}}$. We will be using $\hbar\omega_y$ as the unit of energy. The linear system of equations for the coefficients $a$ and $b$ becomes
\begin{align}
&(\varepsilon_o(n_x, n_y) - \varepsilon) a_{n_x, n_y} + \beta \bigg [  i \sqrt{n_y} b_{n_x, n_y - 1} - i \sqrt{n_y + 1}b_{n_x, n_y + 1}  \nonumber\\&- \sqrt{\gamma n_x} b_{n_x - 1, n_y} + \sqrt{\gamma(n_x + 1)}b_{n_x + 1, n_y}  \bigg ] = 0 \nonumber\\&
(\varepsilon_o(n_x, n_y) - \varepsilon) b_{n_x, n_y} + \beta \bigg [  i \sqrt{n_y} a_{n_x, n_y - 1} - i \sqrt{n_y + 1}a_{n_x, n_y + 1}  \nonumber\\&+ \sqrt{\gamma n_x}a_{n_x - 1, n_y} - \sqrt{\gamma(n_x + 1)}a_{n_x + 1, n_y}  \bigg ] = 0,
\end{align}
where $\varepsilon_o(n_x, n_y) = \gamma(n_x + \frac{1}{2}) + (n_y + \frac{1}{2})$.
This set of equations cannot be solved analytically and one has to resort to numerical methods. In the
symmetric case where $\gamma=1$, one could also have used a basis based on the solution of the 
harmonic oscillator potential in cylindrical coordinates \cite{sinha2011,liu2012,hu2012}. However, 
this is not appropriate in our case since the cylindrical symmetry is broken in the plane for $\gamma\neq 1$
and the problem is therefore better handled using the Cartesian basis expansion presented above.

\begin{figure} 
\begin{center}
\includegraphics[scale=0.30]{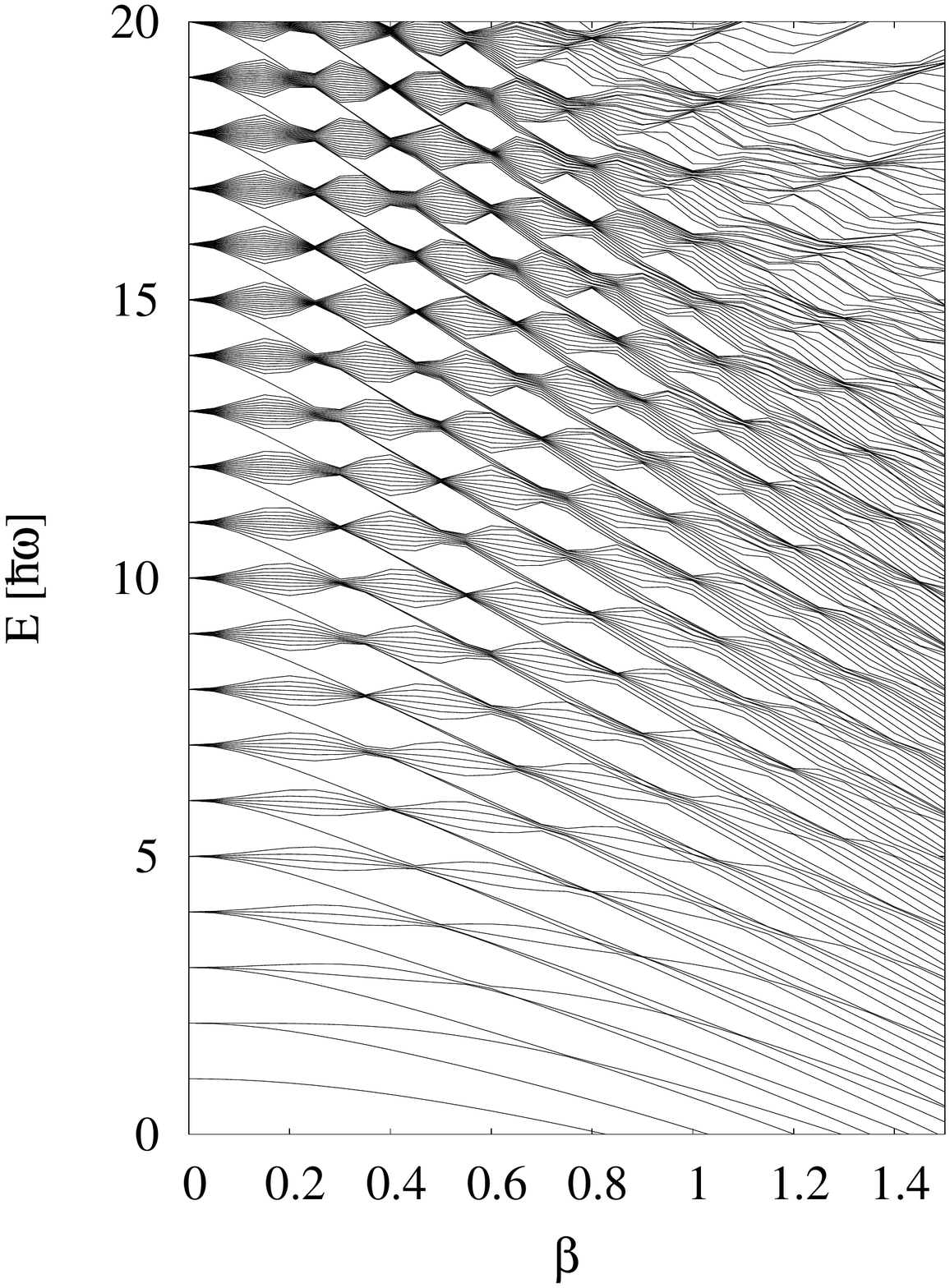}
\includegraphics[scale=0.30]{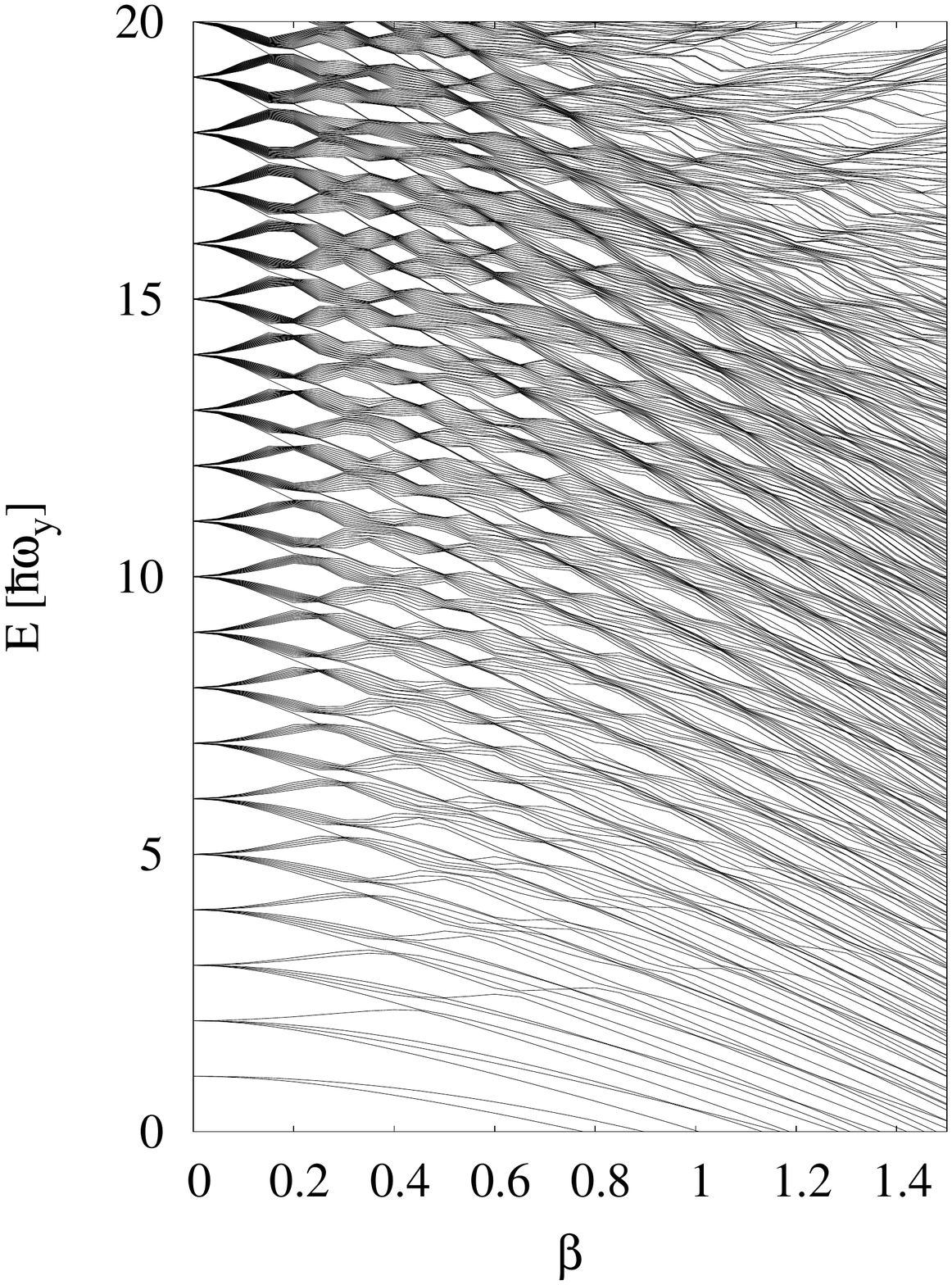}
\end{center}
\caption{Energy as function of the dimensionless spin-orbit coupling parameter $\beta$ for the case of 
equal frequencies $\omega=\omega_x=\omega_y$ with no Zeeman shift (left panel) and including a Zeeman shift
of magnitude $\mu B=\hbar\omega$.}
\label{figure1}
\end{figure}

\section{Single-particle spectra}
We now present the results of our study of the single-particle spectral structure in the presence
of a Rashba spin-orbit term and with an external trap that can be deformed. All the results presented here
are in the regime $0\leq \beta\leq 1.5$. It is a simple matter to go to even higher Rashba couplings
but it requires the use of a bigger single-particle basis in order for all states to properly converge. 
The results we present below have all been obtained by using a basis with {about 700 single-particle states.
However, due to deformation the maximum number of quanta, $\textrm{max}(n_x)$ and $\textrm{max}(n_y)$, 
is generally not equal to each other since we cut the basis in energy space, 
i.e. $\textrm{max}(\varepsilon_0(n_x,n_y))=\gamma\,(\textrm{max}(n_x)+1/2)+(\textrm{max}(n_y)+1/2)$ is 
our cut-off that restricts the values of $n_x$ and $n_y$.

\subsection{Symmetric Trapping Potential}
In the left panel in figure~\ref{figure1} we show the single-particle energy spectrum in a 
two-dimensional spherical ($\gamma = 1$) harmonic oscillator 
as function of the coupling strength, $\beta$. The external magnetic field is not included here ($\mu B = 0$). The oscillator degeneracies that are 
well-known for $\beta=0$ are lifted as $\beta$ increases and the oscillator shells 
become more and more mixed.  Each level is still doubly degenerate due to 
time reversal symmetry mentioned above.
The lowest levels decrease and for sufficiently large strengths the behavior 
approaches a parabolic dependence on $\beta$ that can be found in a 
semi-classical treatment \cite{ghosh2011}. The 
fact that we recover this quadratic behavior for the lowest states imply that 
our calculations capture the large $\beta$ limit for these states. Intuitively,
we can understand the behavior by taking a large $\alpha_R$ limit where we 
neglect all but the Rashba term in the Hamiltonian in \eqref{hamil}. The 
energy should then be proportional to $\alpha_R |\bm p|$, where $\bm p$ 
is the momentum. However, since the only available velocity is $\alpha_R$
itself, we can infer that the energy must scale like $m\alpha_{R}^{2}$.
In section~\ref{perturb} we find this same scaling in perturbation theory.
The perturbation theory below shows that the scaling is $-m\alpha_{R}^{2}$, 
i.e. the energy decreases with $\beta$ in general. This can be understood
again by neglecting the trap in which case the minimum of the free
particle dispersion is at $-m\alpha_{R}^{2}$. This is discussed in 
reference~\cite{vya2011} with an argument that is the same as 
the one we use for the one-dimensional case in section~\ref{oneD} 
(completing the square and obtaining equation~\eqref{1Dspec}).

The many levels are spread out over 
energies and overall they cover rather densely the energy space.  In many 
regions the spectrum resembles a continuous distribution where the individual 
levels cannot be distinguished. The deviations from the regular 
picture seen in the top right-hand corner on both panels in figure~\ref{figure1}
is due to the numerics and the use of a finite basis set. We have numerically 
checked that this can be remedied by systematic expansion of the basis size.

We note the many avoided crossings that can be seen in the spectrum. 
The fact that the Hamiltonian does not preserve the usual spherical 
symmetry (or cylindrical symmetrty in the plane) means that one cannot 
decouple sectors of given angular momentum or parity. Only time-reversal
symmetry remains (in the absence of a Zeeman term). Other studies 
have used cylindrical expansions \cite{liu2012,hu2012} but at the 
expensive of coupling different angular momenta. Here we use the 
Cartesian basis from the start since we find this more convenient. 
The lack of symmetry is reflected in the spectrum which has 
avoided crossings at every point where levels approach each other. 

The key features are found in the low-density pockets 
of the energy spectrum where no levels are found. We see that we have a spectrum 
that can open and close 'super' gaps as a function of the Rashba coupling 
strength. These 
regions are significant since we would expect that a system with this level 
density would tend to be more stable with respect to the application of weak
interactions between the particles or perturbations from the background since 
the larger gaps hinder strong mixing of states. In the opposite case
where the level density is high we expect to see strong mixing and 
small perturbations could have large effects. What is particularly 
remarkable here is the fact that the Rashba term can almost 
cancel the effects of the trap, i.e. the spacing of trap levels at $\beta=0$
of one $\hbar\omega$ is strongly reduced at $\beta\sim 0.2$ for total 
energies above $\sim 5\hbar\omega$. Indeed these sorts of changes are 
reflected on the properties of a many-body system in a symmetric
trap as discussed for the case of condensates in references \cite{sinha2011}, 
\cite{liu2012} and \cite{hu2012}. A recent study has shown that similar 
spectra can be obtained by mapping the current problem to a quantum Rabi 
model \cite{chen2013}.

On the right panel in figure~\ref{figure1} we show again the case of equal 
frequencies, $\omega=\omega_x=\omega_y$, but this time with the inclusion of a
Zeeman field of magnitude $\mu B=\hbar\omega$. The Zeeman field will not influence
the oscillator levels but only displace the spin components. Note that 
the absolute ground state of the spectrum now starts at zero energy and then
decreases. We have opted to keep the same vertical scale on both 
panels in figure~\ref{figure1} in order to make a comparison of 
the overall spectral density at higher energies and therefore 
the Zeeman shifted ground state level cannot be seen.
What we find
is that the Zeeman split will tend to counteract the effect of the 
Rashba term at small $\beta$ so that we now find a spectrum with 
large gaps at $\beta\sim 0.2$. However, since the spin splitting sends
states up and down in the spectrum it tends to generate a sort
of 'two-gap' structure around $\beta\sim 0.2$ with one large and 
one smaller gap, intertwined by a number of densely spaced states.
The conclusion is that both Rashba and Zeeman can be used as an 
experimental handle on the density of states in the single-particle
spectrum as the density and structure can be changed by varying 
one while keeping the other fixed.

\begin{figure} 
\begin{center}
\includegraphics[scale=0.3]{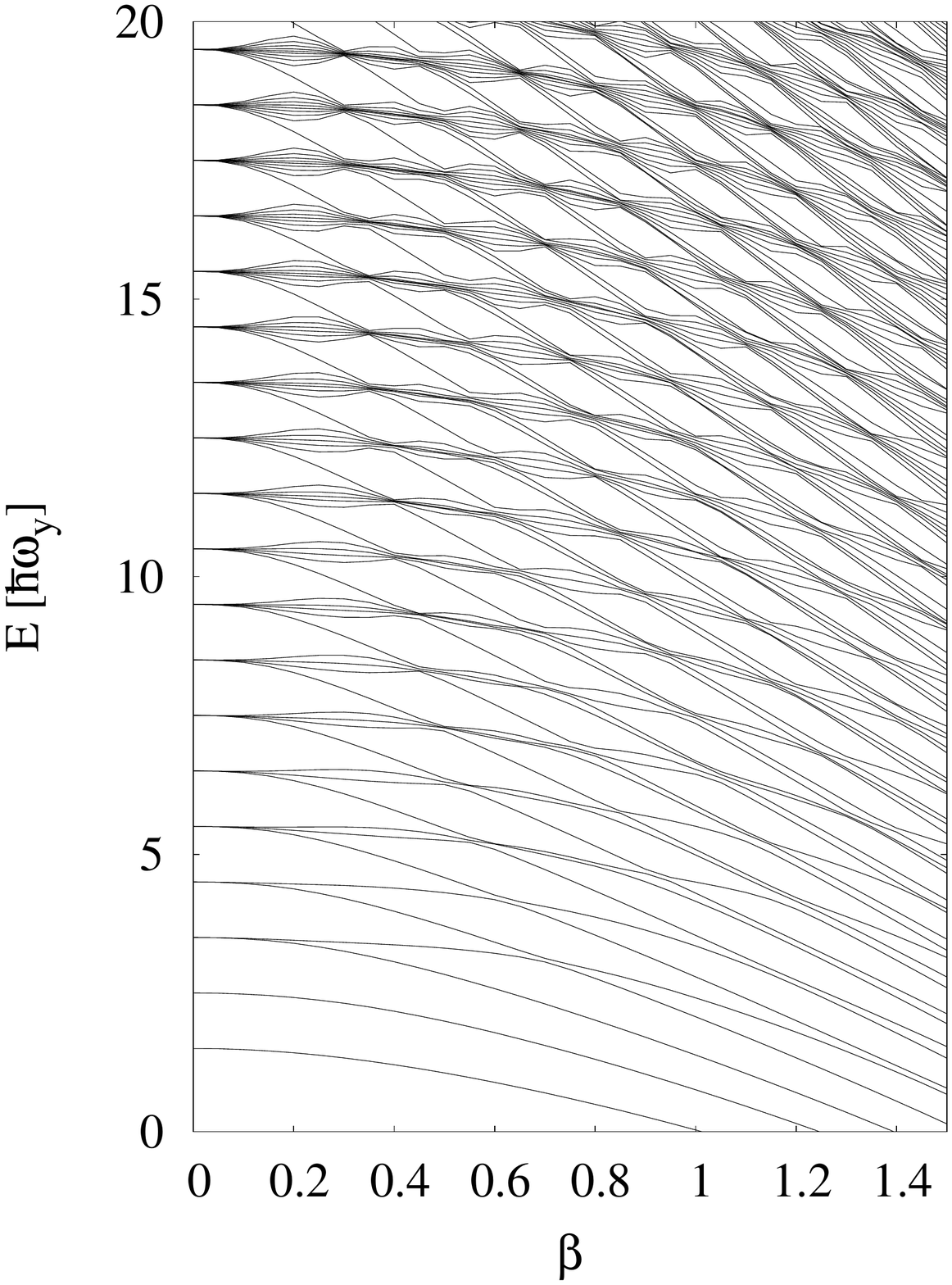}
\includegraphics[scale=0.3]{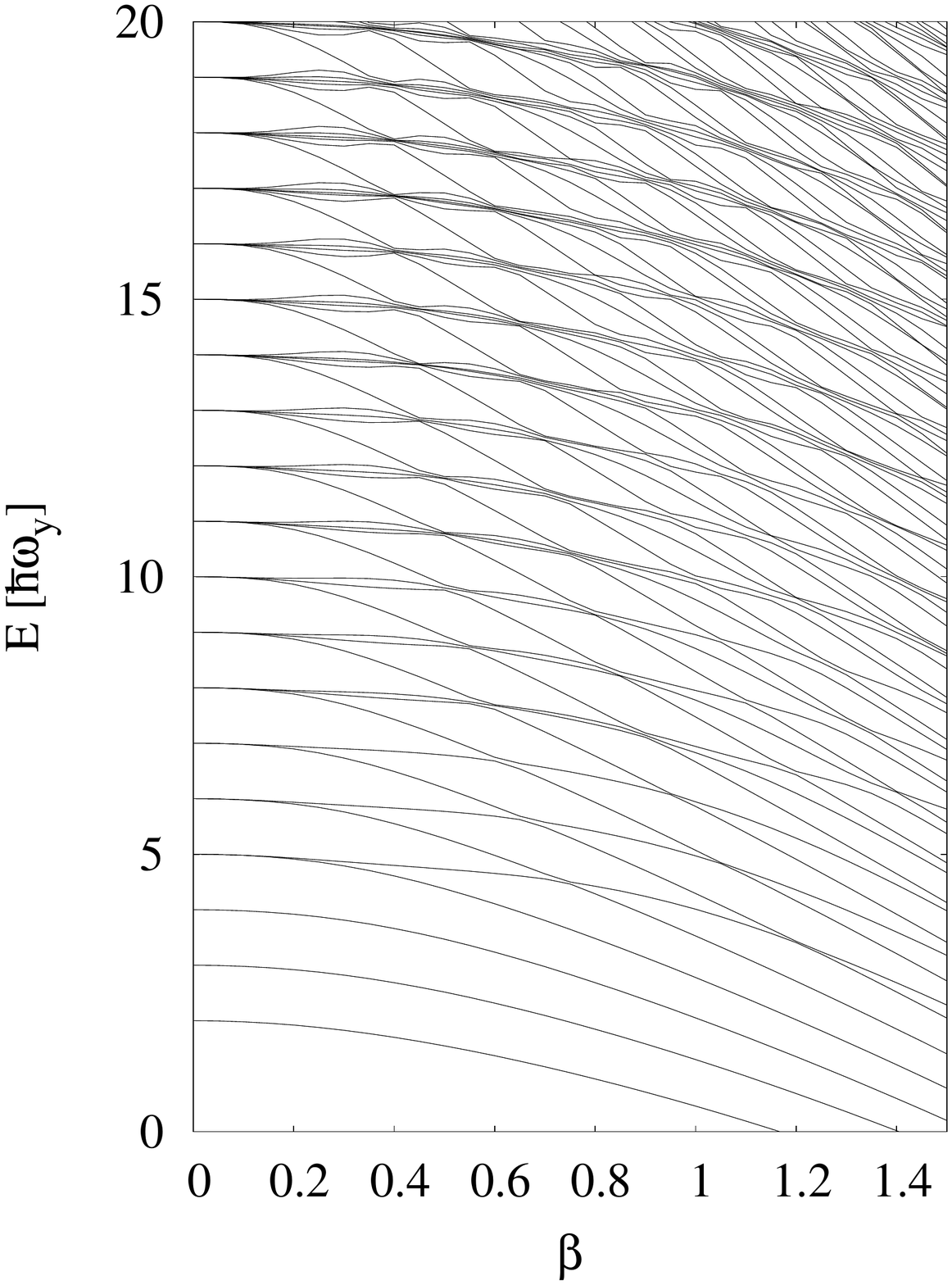}
\end{center}
\caption{Energy as function of dimensionless spin-orbit coupling parameter $\beta$ for the case where the 
oscillator potential is deformed. The left panel has $\gamma=\tfrac{\omega_x}{\omega_y} = 2$ and the 
right panel has $\gamma=3$.}
\label{figure2}
\end{figure}

\subsection{Deformed Trapping Potentials}
We now consider the case of 
an oscillator potential that is non-spherical corresponding to different frequencies 
in the $x$ and $y$ directions. In this situation, high degeneracy is 
found for special frequency ratios of deformed harmonic oscillators. These special 
configurations occur when the frequency ratios equal ratios of small integers 
as $2/1$, $3/2$, etc. This is due to the harmonic oscillator spectrum being 
linear in both the frequencies and quantum numbers in the different spatial directions. 
The harmonic oscillator has especially high degeneracy compared to deformations of 
(almost) all other radial shapes. This means that gaps are more likely
to occur in the spectrum. 
The implication for pure oscillators with no Rashba or Zeeman terms 
is that degeneracies are periodically occurring 
as function of the frequency ratio, and that degeneracies are washed out in between 
the highly degenerate points. 
In figure~\ref{figure2} we show the spectrum for frequency ratios $\gamma=2$ (left panel)
and $\gamma=3$ (right panel) as a function of Rashba coupling strength, $\beta$, with 
no Zeeman field. For these ratios we clearly see that the spectrum is going towards
a more degenerate form with stronger gaps for increasing $\gamma$ in comparison to 
the $\gamma=1$ case shown in figure~\ref{figure1}. Indeed for $\gamma=3$ we have to 
go up to energies of $15\hbar\omega_y$ and above before we can see the same behavior
as for lower energies in the left panel of figure~\ref{figure1}. When including a 
Zeeman term for these ratios (not shown here) we see the same trends as seen 
from left to right in figure~\ref{figure1}, i.e. the spectrum becomes slightly 
more dense overall and there is competition between Rashba and Zeeman terms.

A completely different scenario is displayed in the case where
$\gamma$ is not a ratio of small integers. In figure~\ref{figure3} we
show the single-particle spectrum for $\gamma=1.57$ (left panel) and 
$\gamma=2.17$ (right panel). For these ratios the states are more evenly
distributed and for $\gamma=1.57$ we see an almost constant 
spectral density for energies of $5\hbar\omega_y$ and above, while for 
$\gamma=2.17$ this is not seen until about $10\hbar\omega_y$ and above. 
Comparing the left and right-hand panels in figure~\ref{figure3}, we see
an overall tendency for the larger $\gamma$ to have a smaller overall 
density of levels since this is closer to the one-dimensional limit that 
we will return to momentarily. The overall quadratic decrease with 
$\beta$ is still seen as in figures~\ref{figure1} and \ref{figure2}. 
Comparing the results presented in figure~\ref{figure2} for the 
ratios $\gamma=2$ and $\gamma=3$ to those of figure~\ref{figure3} 
with $\gamma=1.57$ and $\gamma=2.17$ we thus conclude that the 
deformation of the trap is another experimental way to change the 
spectral structure and density. However, for the latter ratios
the influence of the Rashba coupling is diminished somewhat as
the density changes in a much more smooth manner as compared to 
figures~\ref{figure1} and \ref{figure2}. This implies that 
the choice of deformation is very important when studying
the effects of spin-orbit coupling on trapped systems.

\begin{figure} 
\begin{center}
\includegraphics[scale=0.3]{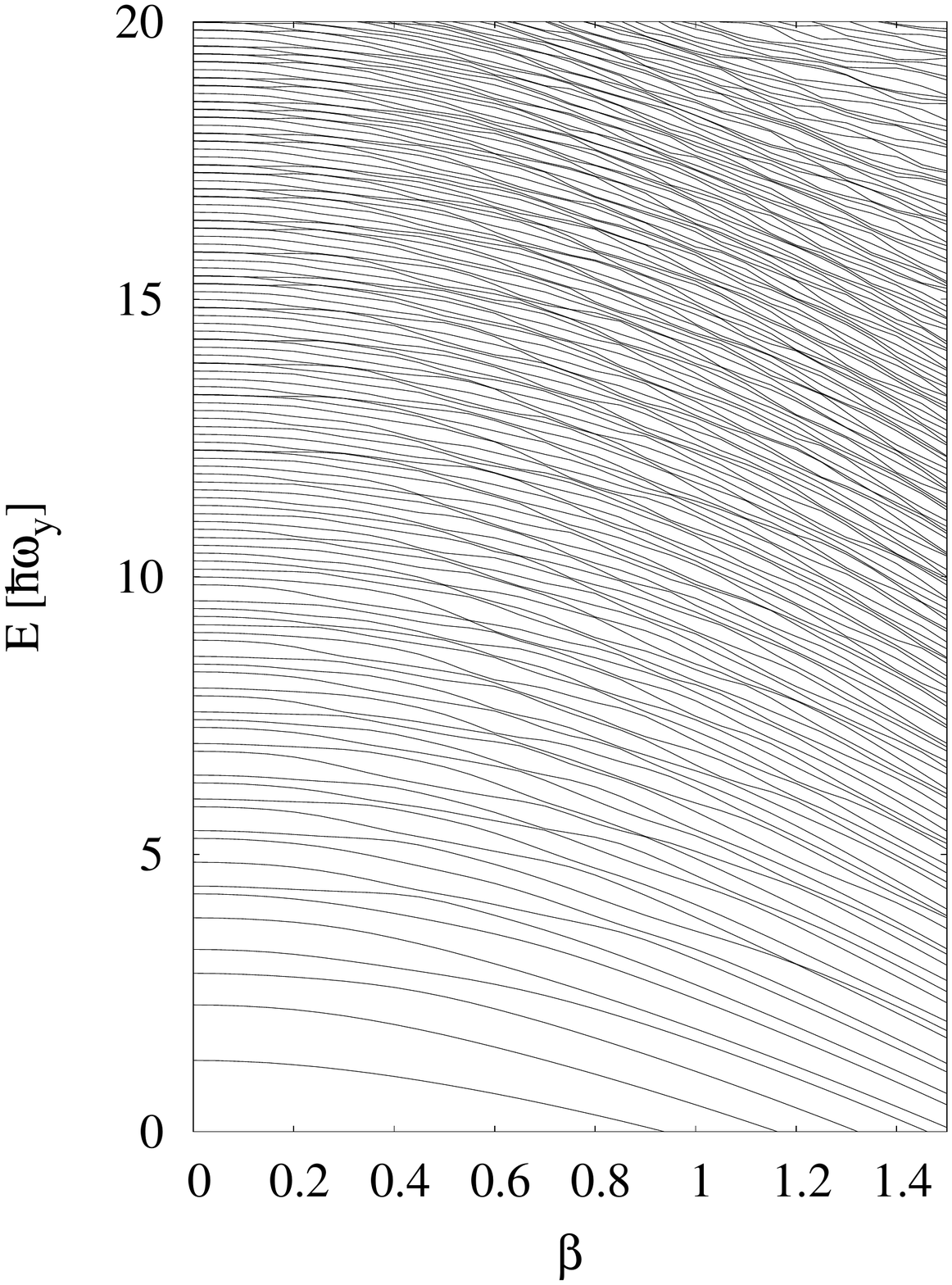}
\includegraphics[scale=0.3]{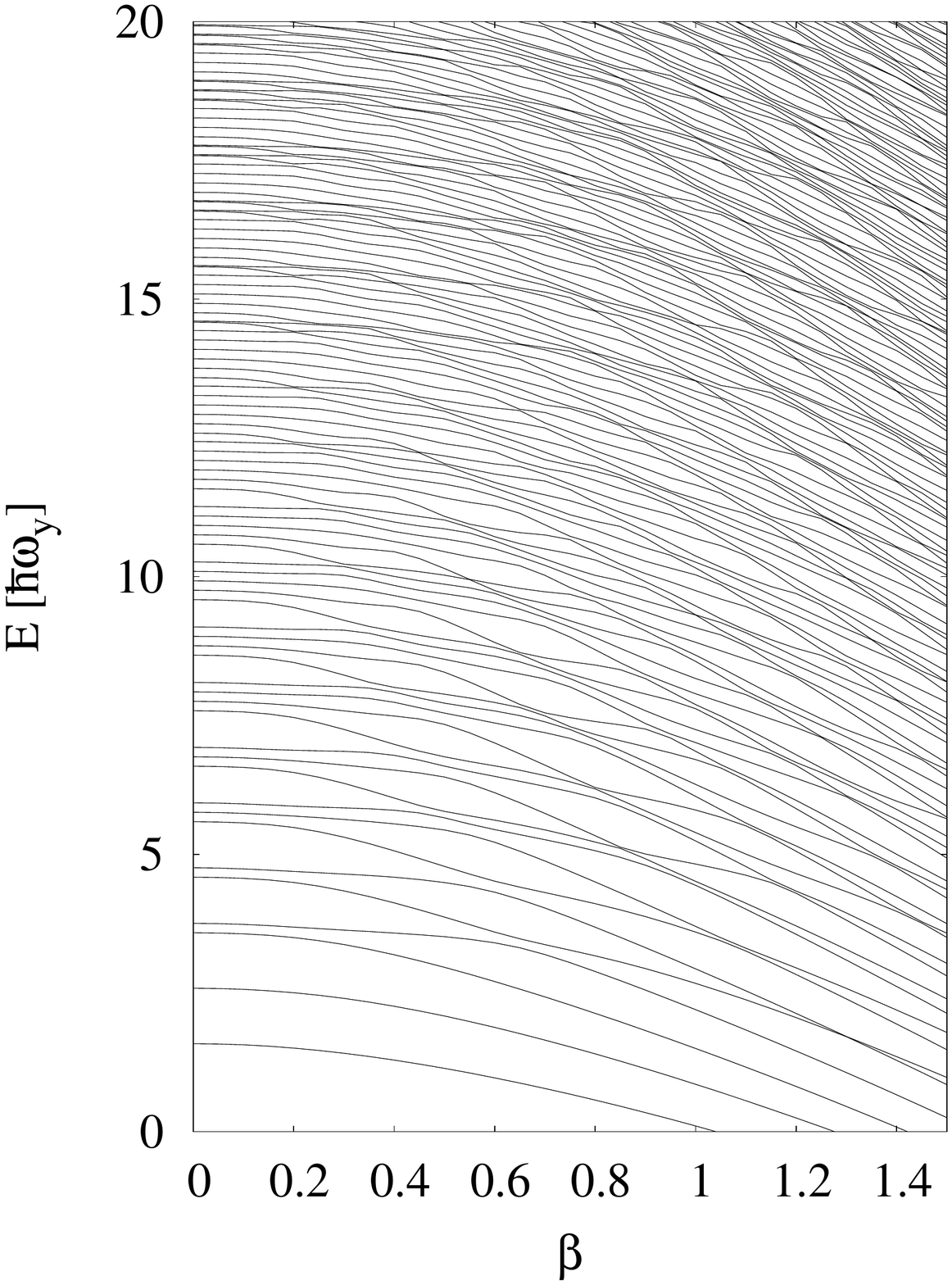}
\end{center}
\caption{Same as figure~\ref{figure2} but with ratios $\gamma=1.57$ in the left panel and $\gamma=2.17$ in the right panel.}
\label{figure3}
\end{figure}

\subsection{One-dimensional Limit}\label{oneD}
Increasing the frequency ratio towards infinity separates the Hamiltonian in a 
low-energy one-dimensional part very weakly coupled to the high-energy one-dimensional 
part in the other direction. With our conventions, the limit $\gamma\gg 1$ corresponds to 
a strongly confined motion along the $x$-direction and a shallow confinement along the 
$y$-direction. These decoupled one-dimensional equations can be 
solved analytically with the corresponding Rashba couplings. The Pauli 
matrix for either $\sigma_x$ or $\sigma_y$ implies that one should use a basis of eigenfunctions 
for the operator with the small frequency, i.e. $\omega_y$, since the
other direction has a very large oscillator frequency that effectively freezes the 
motion.
This linear combination of equal amplitude 
spin-up and spin-down spinors immediately decouple the equations of motion into 
separate oscillator eigenvalue problems. The eigenvectors of $\sigma_x$ are
\begin{equation}
\Phi_{\pm} =
\begin{pmatrix}
1\\
\pm 1
\end{pmatrix},
\end{equation}
and these can now be used as the two-component basis instead of the spin-up and spin-down spinors.
The one-dimensional Schr{\"o}dinger equation then becomes
\begin{equation}
\bigg (\frac{p_y^2}{2m} + \frac{1}{2}m\omega_y^2 y^2 \pm \alpha_R p_y - E\bigg ) \Phi_{\pm} f_\pm(y)= 0,
\end{equation}
where $f_\pm(y)$ is the radial wave function.
We rewrite these equations as 
\begin{equation}
\bigg [\frac{1}{2m} (p_y \pm m \alpha_R)^2 + \frac{1}{2}m\omega_y^2 y^2 - \frac{m \alpha_R^2}{2} - E \bigg ] \Phi_{\pm}f_\pm(y) = 0,
\end{equation}
from which the harmonic oscillator solutions can be directly inferred to be
\begin{equation}\label{1Dspec}
E_{n_y} = - \frac{1}{2} m \alpha_R^2 + \hbar \omega_y (n_y + \frac{1}{2})
\end{equation}
The first term is simply the quadratic decrease of the energy with $\beta$ 
that we already noted above. It can be interpreted as a Galilean boost
by the Rashba velocity, $\alpha_R$.
We can compare this dispersion to the 
one-dimensional solution with no external trapping potential in the 
$y$-direction which is simply
\begin{equation}\label{1Dhomo}
E_{k_y}=\frac{\hbar^2k_{y}^{2}}{2m}\pm \alpha_R \hbar k_y,
\end{equation}
where $k_y$ is the one-dimensional momentum along the $y$-direction. These two 
dispersion relations look very different but can be reconciled by recalling that the 
virial theorem tells us that $\langle \tilde{p}_{y}^{2}\rangle\propto m\hbar\omega_y n_y$
where $\tilde{p}_y=p_y\pm m\alpha_R$. From the second term in equation~\ref{1Dspec} we
would thus get linear, $\alpha_R\langle p_y\rangle$, and quadratic, $\langle p_{y}^{2}\rangle$, 
dependence on the momentum. This is the same structure as equation~\ref{1Dhomo} and
we see that the harmonically trapped system behaves similarly to the free-space
case when considering the expectation value of the momentum.

In figure \ref{figure4} we show the spectral structure as one approaches the 
one-dimensional limit. In the left panel of figure~\ref{figure4} we have 
$\gamma=5$ and in the right panel $\gamma=10$. 
The low-energy eigenvalues reduce to the equidistant harmonic oscillator 
spectrum with the same frequency but shifted quadratically
with the spin-orbit coupling strength.  The remarkably simple emerging
feature is that the lowest part of the spectrum for an even modest deformation 
already resembles the equidistant one-dimensional limit. This can be 
clearly seen by making a comparison of figure~\ref{figure4} to 
figures~\ref{figure2} and \ref{figure3}.
As the deformation increases an increasing part of the low-energy spectrum 
approaches the one-dimensional limit. 
This feature can be understood from the weak coupling of two oscillators 
with very different frequencies. The perturbation on the
lowest energy states in the spectrum from the lowest of the large frequency ($\omega_x$) 
states is proportional 
to the square of the coupling strength divided by the
energy difference by second order perturbation theory. 
This tells us that the one-dimensional limit is approached 
in the bottom 
of the spectrum with increasing deformation. This low-energy spectrum is much 
simpler and much less dense than that of the two-dimension spherical oscillator.
Note that the spectrum is still two-fold degenerate due to the time-reversal 
symmetry, but now it is practically back to the standard equidistant scheme
typical of harmonic confinement.

\begin{figure} 
\begin{center}
\includegraphics[scale=0.3]{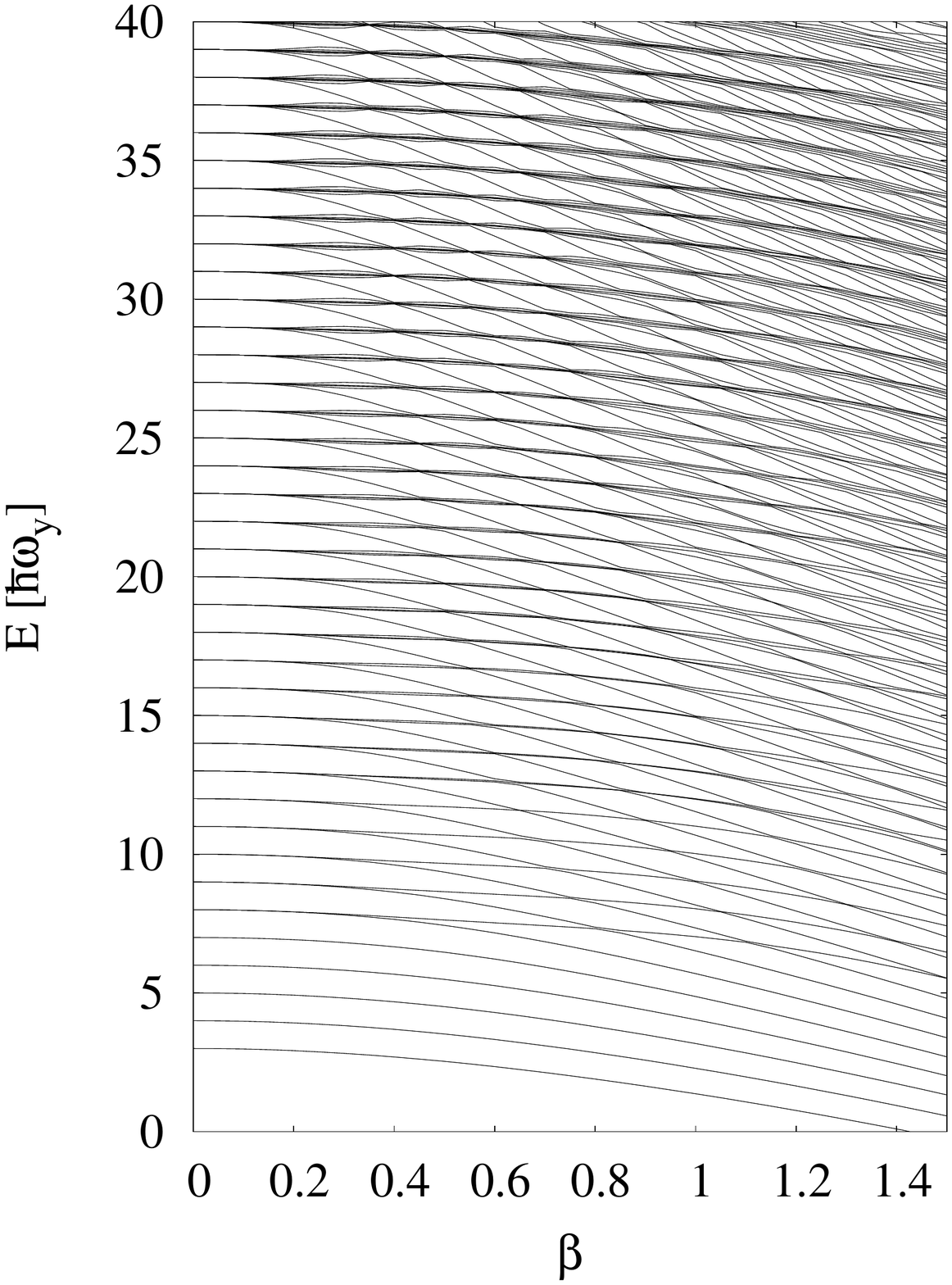}
\includegraphics[scale=0.3]{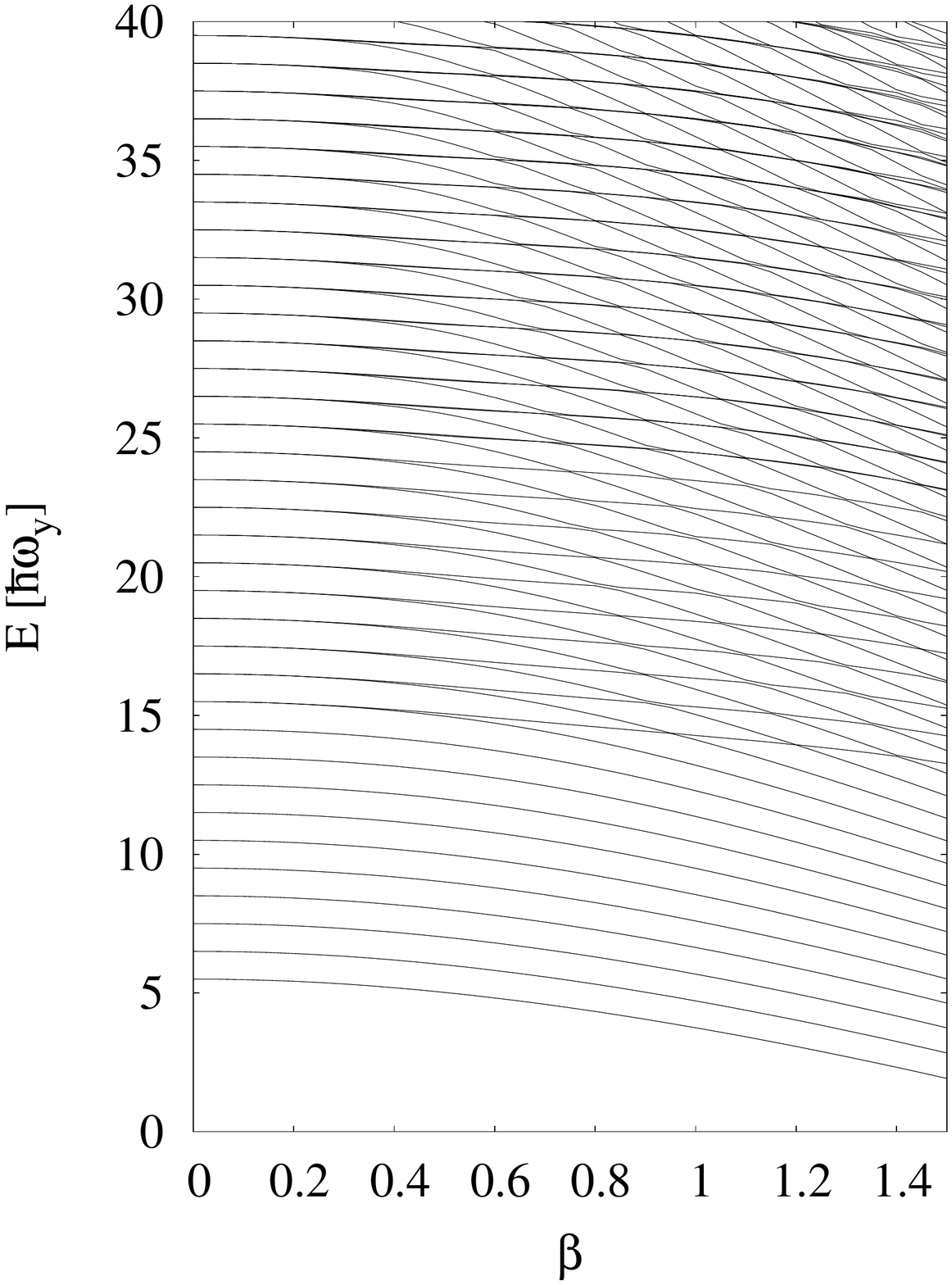}
\end{center}
\caption{Same as figure~\ref{figure2} and \ref{figure3} for $\gamma=5$ (left panel) and $\gamma=10$ (right panel). These 
results approach the limit of an effectively one-dimensional system, i.e. $\gamma\gg 1$.}
\label{figure4}
\end{figure}

\subsection{Perturbation theory for weak spin-orbit coupling}\label{perturb}
The weak coupling limit, $\beta \to 0$, can be investigated in detail using perturbation theory. 
In the absence of deformation and external magnetic field, the system is highly degenerate (Kramers and oscillator degeneracies) and 
requires degenerate perturbation theory to second order (first order vanishes as we discuss momentarily).
The easiest way to avoid the complicated expressions of the degenerate formalism is to consider a deformed oscillator trap with an external magnetic field which lifts all degeneracies. By continuity, we get the non-deformed and zero external field by taking this limit
at the end of the calculation.

The Hamiltonian is $\hat H = \hat H_0 + \hat V_R$, where 
\begin{equation}
\hat H_0 = \frac{{\hat p}_x^2}{2m} + \frac{1}{2} m \omega_x^2 x^2 + \frac{{\hat p}_y^2}{2m} + \frac{1}{2} m \omega_y^2 y^2 - \mu B \hat \sigma_z
\end{equation}
is the unperturbed part of the Hamiltonian and the perturbation is
\begin{equation}
 \hat V_R = \alpha_R (\hat \sigma_x \hat p_y - \hat \sigma_y \hat p_x).
\end{equation}
Then the Schr{\"o}dinger equation for the unperturbed Hamiltonian is
\begin{equation}
\hat H_0 \psi^{(0)}(x, y, \sigma) = E^{(0)} \psi^{(0)}(x, y, \sigma),
\end{equation}
with the well-known solutions: $\psi^{(0)}(x, y, \uparrow) = \phi_{n_x, n_y}(x, y)\bigl(\begin{smallmatrix}
1 \\ 0
\end{smallmatrix}\bigr)$ and $\psi^{(0)}(x, y, \downarrow) = \phi_{n_x, n_y}(x, y)\bigl(\begin{smallmatrix}
0 \\ 1
\end{smallmatrix}\bigr)$, with eigenenergies $E^{(0)}_{n_x, n_y, \sigma} = \hbar \omega_x (n_x + \frac{1}{2}) + \hbar \omega_y (n_y + \frac{1}{2}) - \sigma \mu B$, where $\sigma = \pm 1$ for spin up and down, respectively. The spatial parts $\phi_{n_x, n_y}(x, y)$ are 2D harmonic oscillator solutions.

Now we can directly find corrections to the energy. One can see that the diagonal matrix element for $V_R$ is zero, thus the first-order correction will be zero as well. The second-order correction is\cite{landau1977}
\begin{equation}
E^{(2)}_{n_x, n_y, \sigma} = \sideset{}{'}\sum_{m_x, m_y, \sigma'} \frac{|\langle n_x, n_y, \sigma |\hat V_R | m_x, m_y, \sigma' \rangle|^2}{E^{(0)}_{n_x, n_y, \sigma} - E^{(0)}_{m_x, m_y, \sigma'}}.
\end{equation}
We can explicitly write down the matrix elements. Due to the properties of Pauli matrices $\hat \sigma_x$ and $\hat \sigma_y$ only the matrix elements with different spin projections will contribute in the sum.
\begin{align}
&\langle n_x, n_y, \uparrow |\hat V_R | m_x, m_y, \downarrow \rangle = -\alpha_R \sqrt {\frac{m \hbar}{2}} [ \sqrt{\omega_x}(\sqrt{m_x + 1}\delta_{n_x, m_x + 1} -\nonumber\\& - \sqrt{m_x} \delta_{n_x, m_x - 1}) \delta_{n_y, m_y} - i \sqrt{\omega_y}(\sqrt{m_y + 1}\delta_{n_y, m_y + 1} - \sqrt{m_y} \delta_{n_y, m_y - 1}) \delta_{n_x, m_x}]
\end{align}
and
\begin{align}
&\langle n_x, n_y, \downarrow |\hat V_R | m_x, m_y, \uparrow \rangle = \alpha_R \sqrt {\frac{m \hbar}{2}} [ \sqrt{\omega_x}(\sqrt{m_x + 1}\delta_{n_x, m_x + 1} -\nonumber\\& - \sqrt{m_x} \delta_{n_x, m_x - 1}) \delta_{n_y, m_y} + i \sqrt{\omega_y}(\sqrt{m_y + 1}\delta_{n_y, m_y + 1} - \sqrt{m_y} \delta_{n_y, m_y - 1}) \delta_{n_x, m_x}].
\end{align}
We see that only transitions between the nearest states contribute to the sum. It is straightforward now to write down the second-order corrections for the ground state energy
\begin{equation}
E^{(2)}_{0, 0, \uparrow} = - \frac{m \alpha_{R}^2}{2} \big [ \frac{\hbar \omega_x}{\hbar \omega_x + 2\mu B} + \frac{\hbar \omega_y}{\hbar \omega_y + 2\mu B}\big ].
\end{equation}
For a case of $\mu B \ll \hbar \omega_y$ we get a simple term, which appears natural considering that the dimension of $\alpha_R$ is the velocity
\begin{equation}
E^{(2)}_{0, 0, \uparrow} = -m \alpha_{R}^2.
\end{equation}
This equation is valid for coupling parameter values smaller than the oscillator velocity, $\alpha_R \ll \sqrt \frac{\hbar \omega}{m}$. For a regime of larger $\alpha_R$ the contribution of all states becomes crucial, therefore one has to consider the higher-order perturbation corrections to energy. However, for the case of $\alpha_R \gtrsim \sqrt \frac{\hbar \omega_y}{m}$ the behavior of the ground state energy is still quadratic in $\alpha_R$~\cite{hu2012}.

\section{Average spectral properties}
The single-particle spectra reveal regions of stability for specific finite particle numbers.
The stability is closely related to the density of states at the Fermi level. To examine these properties for finite systems, we first calculate the average behavior of the density of states.
Subsequently we relate to the critical strength for superfluidity which becomes strongly dependent on the spectral properties.

\subsection{Density of states}
The density of single-particle states $g(\varepsilon)$ is a sum of delta functions for a system with a discrete energy spectrum.  To emphasize the underlying structures we broaden each of the eigenvalues, $\varepsilon_i$, by a normalized gaussian. Then we get the average density of states,
\begin{equation}
g_{\delta}(\varepsilon) = \frac{1}{\delta \sqrt{\pi}}\sum_{i}(\frac{3}{2} - (\frac{\varepsilon - \varepsilon_i}{\delta})^2) \exp{[-(\frac{\varepsilon - \varepsilon_i}{\delta})^2]},
\end{equation}
where the second order polynomial is introduced to guarantee that a smoothly varying set of eigenenergies produce the same smooth behavior \cite{brack1972}.

\begin{figure}
\centering
\includegraphics[scale=0.75]{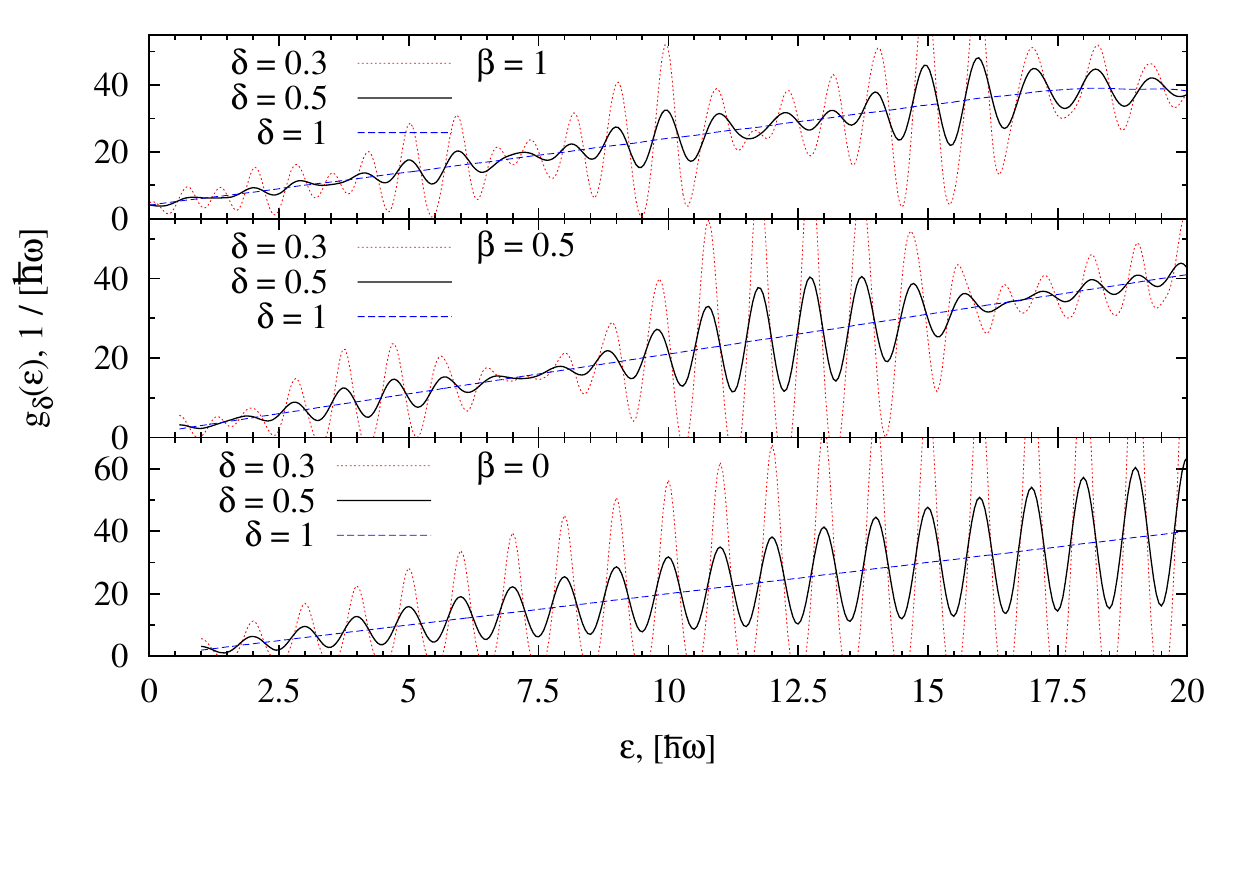}
\includegraphics[scale=0.75]{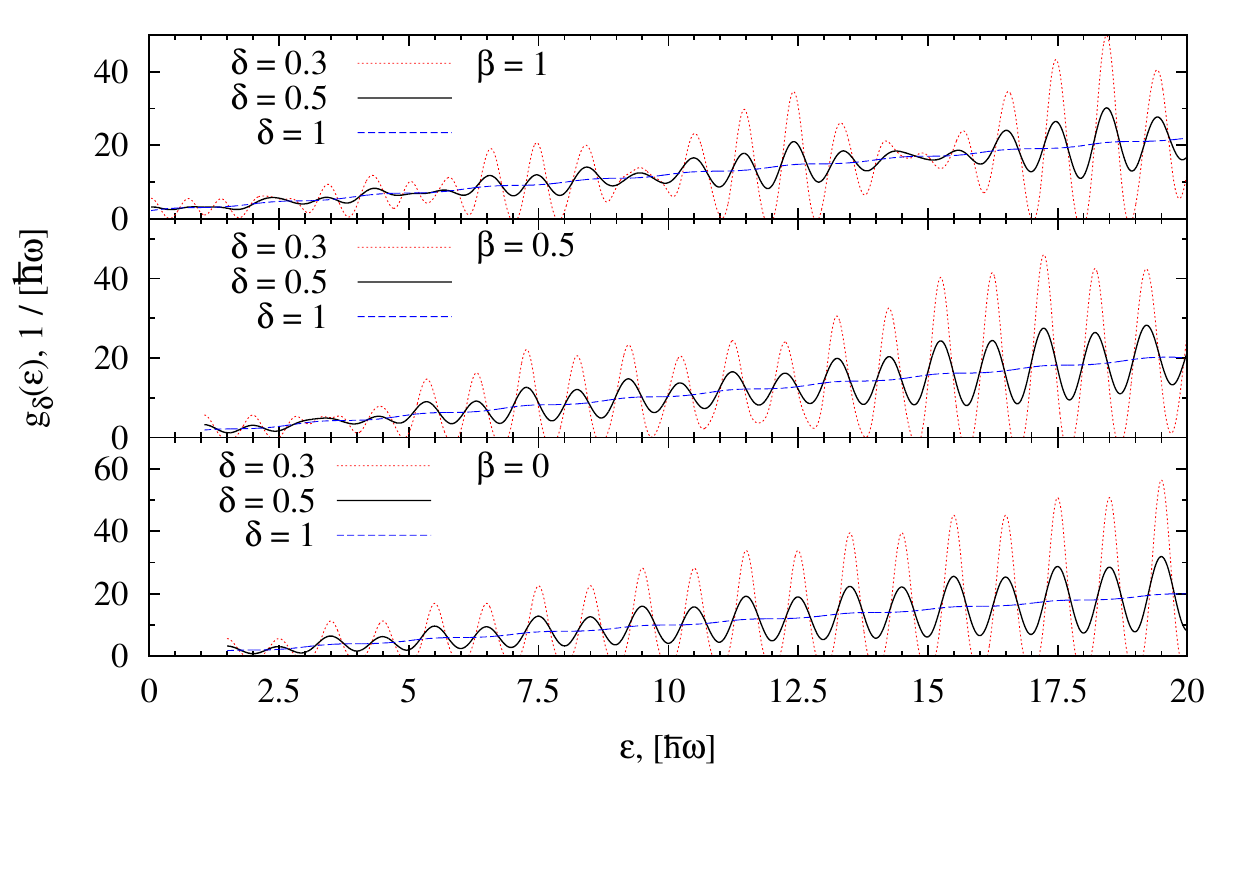}
\includegraphics[scale=0.75]{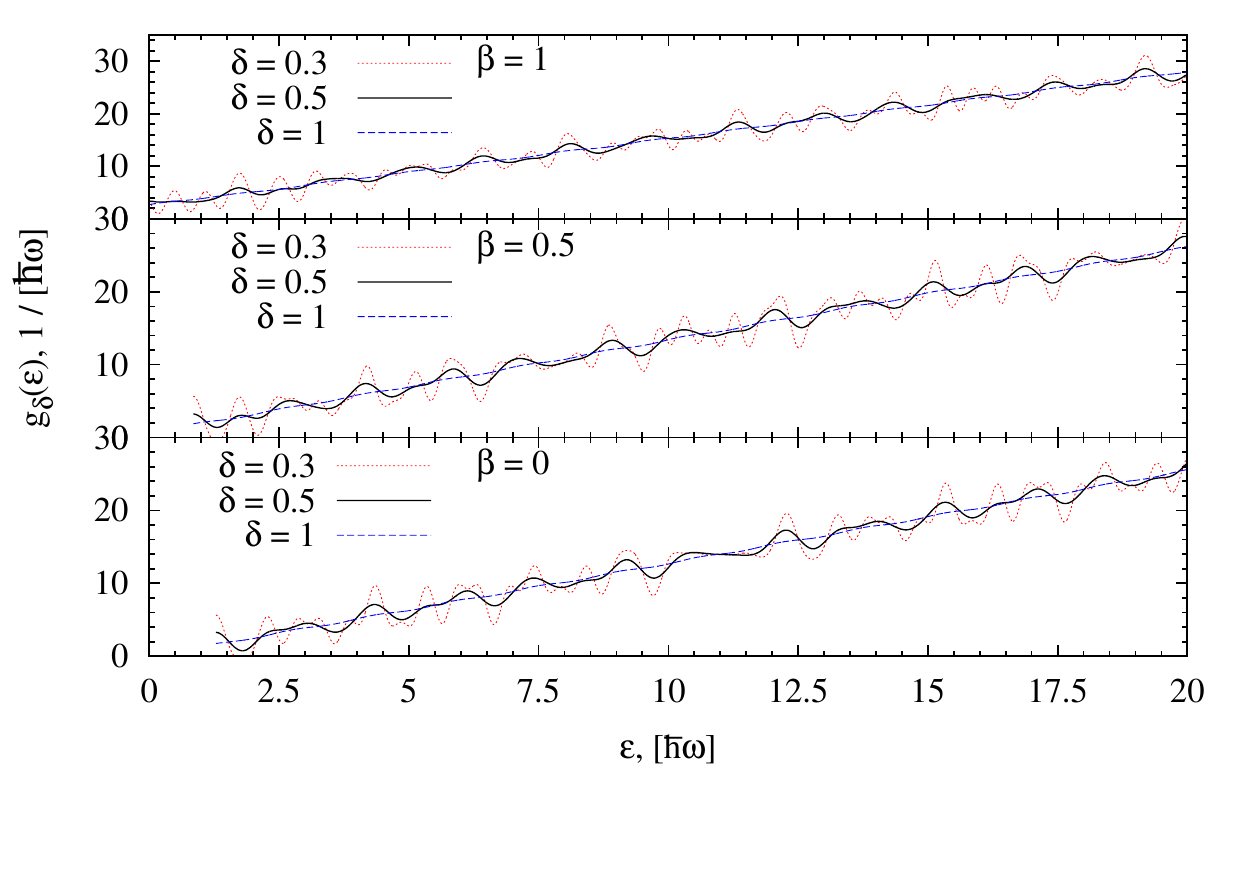}
\caption{Single-particle density of states as function of energy for the spherical $\gamma=1$ case (upper panel) and deformed cases with the frequency ratios $\gamma = 2$ (middle panel) and $\gamma = 1.57$ (lower panel). The smearing parameter, $\delta$, and the dimensionless Rashba coupling paremeters, $\beta$, are given in the panels. Here we set $\omega=\omega_y$.}
\label{figure5}
\end{figure}

If the smearing parameter, $\delta$, is equal to or larger than the shell spacing of the spherical oscillator only average properties are left in $g_{\delta}(\varepsilon)$. By 'shell' we mean the set of states, which corresponds to one oscillator level for the $\beta = 0$ case. To exhibit shell structure $\delta$ has to be less than about half of the shell spacing.
For a spherical two-dimensional oscillator the density of states $g_{\delta}(\varepsilon)$ for $\delta \gtrsim \hbar \omega$ becomes independent of the smearing parameter. In this case the index $\delta$ can be omitted
\begin{equation}
g_{\delta}(\varepsilon) \xrightarrow[\delta \gtrsim \hbar \omega]{} \tilde{g}(\varepsilon) = \frac{2\varepsilon}{(\hbar \omega)^2},
\end{equation}
which is the (semi-classical) density of states of a two-dimensional harmonic oscillator.
The factor of $2$ appears due to the Kramers theorem for a spin-$\frac{1}{2}$ particle. More generally, for a deformed oscillator \cite{pethick2004}
\begin{equation}
\tilde {g}(\varepsilon) \approx \frac{2\varepsilon}{\hbar^2 \omega_x \omega_y} = \frac{2\varepsilon}{\gamma (\hbar \omega_y)^2}
\end{equation}
We can then relate the particle number, $N$, and the Fermi energy, $\varepsilon_F$, which is the energy of the highest occupied state, by $N = \int\limits_0^{\varepsilon_F} \tilde g(\varepsilon) d\varepsilon$.
This gives
\begin{equation}
N \approx \frac{2 \varepsilon_F^2}{2\hbar^2\omega_x \omega_y}
\end{equation}
or by inversion
\begin{equation}
\label{mu_and_N}
\varepsilon_F \approx \hbar \sqrt{N\omega_x \omega_y}.
\end{equation}
For zero temperature, the chemical potential, $\mu$, is equal to the Fermi energy for $N$ particles.

In figure~\ref{figure5}, we show $g_{\delta}(\varepsilon)$ for a number of $\beta$-values and deformations of the external field. The regular behavior is seen for a spherical oscillator for $\beta = 0$.  The amplitudes of the oscillations decrease as the smearing parameter, $\delta$, increases.  When $\delta$ approaches zero the original discrete spectrum is recovered, and when $\delta $ is larger
than $\hbar \omega$ only $\tilde{g}(\varepsilon)$, the smooth (linear) average behavior remains. This regularity changes when $\beta$ assumes non-zero values as seen in the middle and upper panels of figure~\ref{figure5}.  The amplitudes of the oscillations still decrease with increasing $\delta$.  
However, these amplitudes are now irregular functions of $\varepsilon$. Small amplitudes produce the smooth average background behavior for moderate $\delta$-values, and correspond to a dense single-particle spectrum. The opposite applies to large amplitudes where the single-particle structure is more pronounced corresponding to a dilute spectrum. This structural variation changes as function of energy. The other deformation in figure~\ref{figure5} reveals the same overall qualitative behavior, that is regular behavior for $\beta=0$ and variation of the amplitudes as function of $\varepsilon$ for non-zero $\beta$-values. The correspondence between large-small amplitudes and white-black regions are clearly seen by comparing figure~\ref{figure5} and figure~\ref{figure1}. 

The similarities between results of a spherical shape and a small integer ratio of the directional frequencies, $\frac{\omega_x}{\omega_y} = 2$, are due to large degeneracies arising from many coinciding levels. Irrational frequency ratios remove this regularity in the density of states even for $\beta=0$. The most direct implication of small (large) amplitudes at specific energies is low (high) stability for a particle number corresponding to a Fermi level at that energy \cite{brack1972}.
The calculated spectra are single-particle energies for a system of non-interacting fermions. However, similar qualitative behavior would be present for any mean-field approximation of interacting fermions. For a finite number of particles, the spectral properties around the Fermi energy are decisive for corrections beyond the mean-field structure.

\subsection{Critical strength}
Superfluid behavior is possible in the presence of a small residual interaction. For a system of a finite number of particles the microscopic properties of the eigenvalue spectrum determine the minimum strength, $G_c$, of the attractive interaction necessary for creation of a superfluid state
in a fermionic system as we now discuss.

To extract the overall behavior we assume the residual pairing interaction
\begin{equation}
H_R = -\sum_{i, k} G_{ik} {a_k}^{\dagger} {a_{ \overline{k} } }^{\dagger} a_{\overline{i}} a_{i} ,
\end{equation}
where $(a_{k}, a^{\dagger}_{k})$ are annihilation and creation operators for single-particle states $|k \rangle$. The time reversed state of $|k \rangle$ is denoted $|\overline{k} \rangle$. The BCS approximation provides the equations which determine the gaps
\begin{equation}
2 \Delta_k = \sum_{i} \frac{\Delta_i G_{ik}}{\sqrt{(\varepsilon_i - \mu)^2 + {\Delta_i}^2}}.
\end{equation}
The states that provide the largest contribution for pairing solutions are those in the interval, $\pm S$, around the chemical potential, $\mu$, which is equal to the Fermi energy for zero temperature.
For a constant pairing matrix element, $G$, the gap, $\Delta_i$, is also a constant, $\Delta$, which is determined by the states around $\mu$ \cite{leggett2006}. 
The magnitude of $\Delta$ is then determined by 
\begin{equation}
\label{superfluid_gap}
2 =  G \sum_{i} \frac{1}{\sqrt{(\varepsilon_i - \mu)^2 + {\Delta}^2}},
\end{equation}
where the summation now only extends over states with $|\varepsilon_i - \mu| < S$. However, to get a superfluid solution with finite $\Delta$ we must have $G > G_c$, where the critical strength $G_c$ is given by
\begin{equation}
\label{crit_coup}
\frac{2}{G_c} = \sum_{i}\frac{1}{|\varepsilon_i - \mu|}.
\end{equation}
This critical strength arises only due to the finite number of particles. When $N \to \infty$, we have $G_c \to 0$.

For small temperature, the value of $\mu$ is almost inevitably between the last occupied, $\varepsilon_{occ}$, and the first unoccupied level, $\varepsilon_{un}$. It is then a reasonable approximation to choose $\mu = \frac{1}{2}(\varepsilon_{occ} + \varepsilon_{un})$ which depends on the particle number. As $\mu$ increases in steps corresponding to increase of the single-particle levels $(\varepsilon_{occ}, \varepsilon_{un})$ we can for a single-particle spectrum (specified by fixing $\beta$ and $\gamma$) compute $G_c$ as function of $\mu$, or equivalently particle number (through
the usual BCS number equation \cite{pethick2004}).

The approximations to Eqs.~\eqref{superfluid_gap} and \eqref{crit_coup} are quickly formed by substituting the summation by an integral. We immediately get
\begin{equation}
\frac{1}{G_c} \simeq g_{\delta}(\mu) \ln(\frac{2S}{\Delta}),
\label{sf_coup_int}
\end{equation}
\begin{equation}
\frac{1}{G_c} \simeq g_{\delta}(\mu) \ln(\frac{2S}{\varepsilon_{un} - \varepsilon_{occ}}) \simeq g_{\delta}(\mu)\ln(2S g_{\delta}(\mu)),
\label{crit_coup_int}
\end{equation}
where we used that $(\varepsilon_{un} - \varepsilon_{occ})g_{\delta}(\mu) \simeq 1$, following from the basic definition of the doubly degenerate single-particle density of states.

By working with constant matrix elements and critical values, $G_c$, we only include effects of the appropriate residual interaction. 
In particular, we have ignored the reduction due to the fact that the interaction in cold atoms for two-component fermions is 
in the spin singlet channel \cite{pethick2004}, while the spin-orbit coupling mixes triplet and singlet states \cite{zhai2012}. We thus
expect that there will be a level dependent reduction factor (projection onto the spin-singlet states). 
Furthemore, we have regularized the interaction strength by allowing only coupling in the chosen interval around the Fermi level. This 
avoid the more cumbersome regularization procedure required when the interaction acts not only at the Fermi energy but
everywhere \cite{pethick2004}.

\begin{figure}
\centering
\includegraphics[scale=0.75]{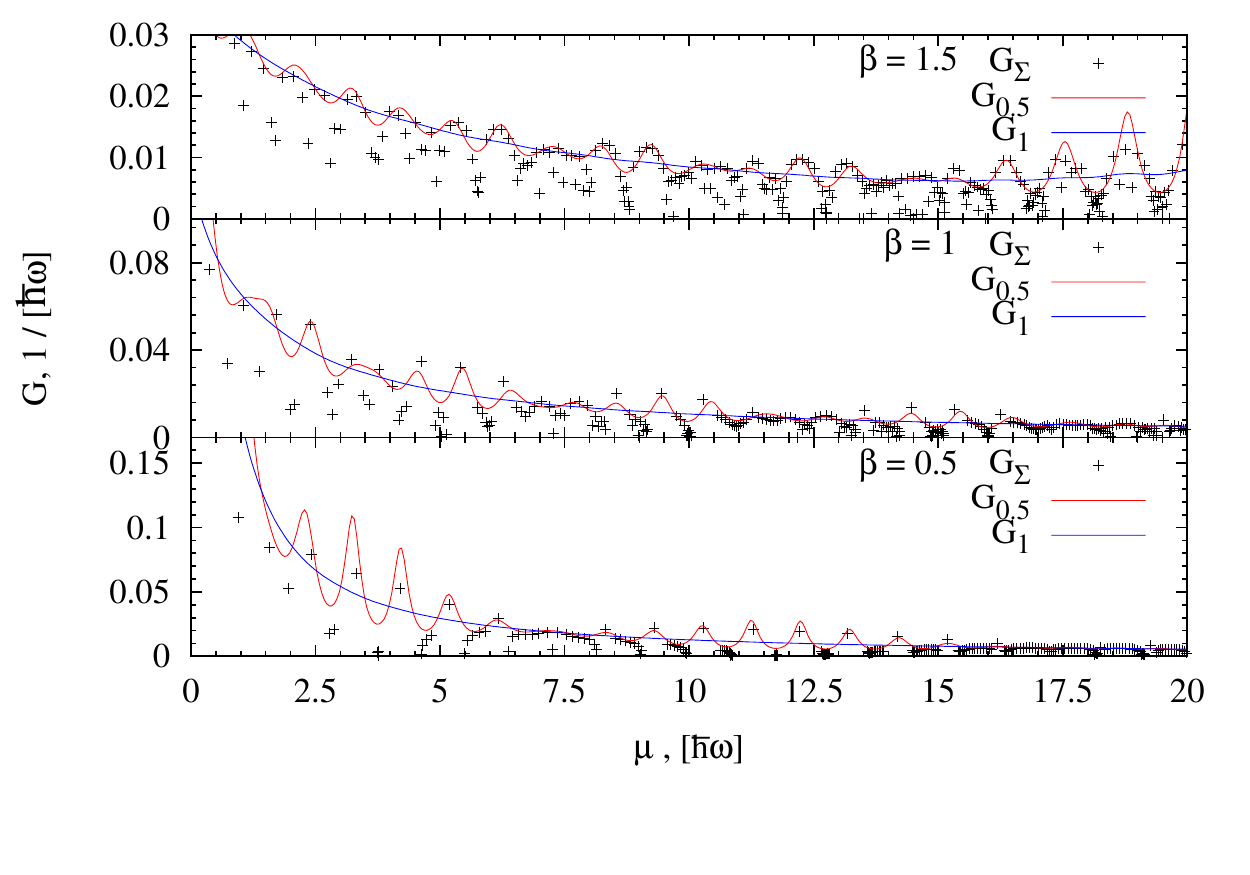}
\includegraphics[scale=0.75]{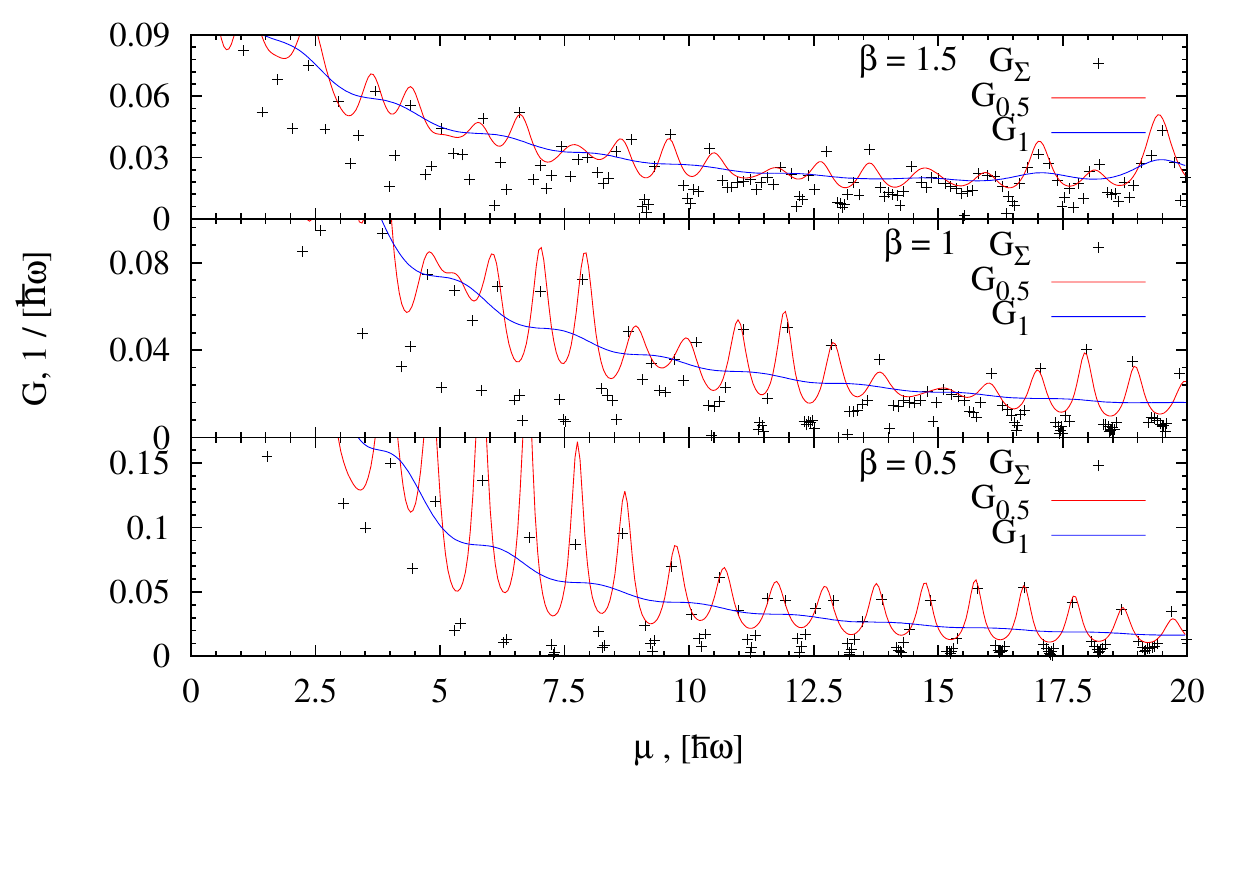}
\includegraphics[scale=0.75]{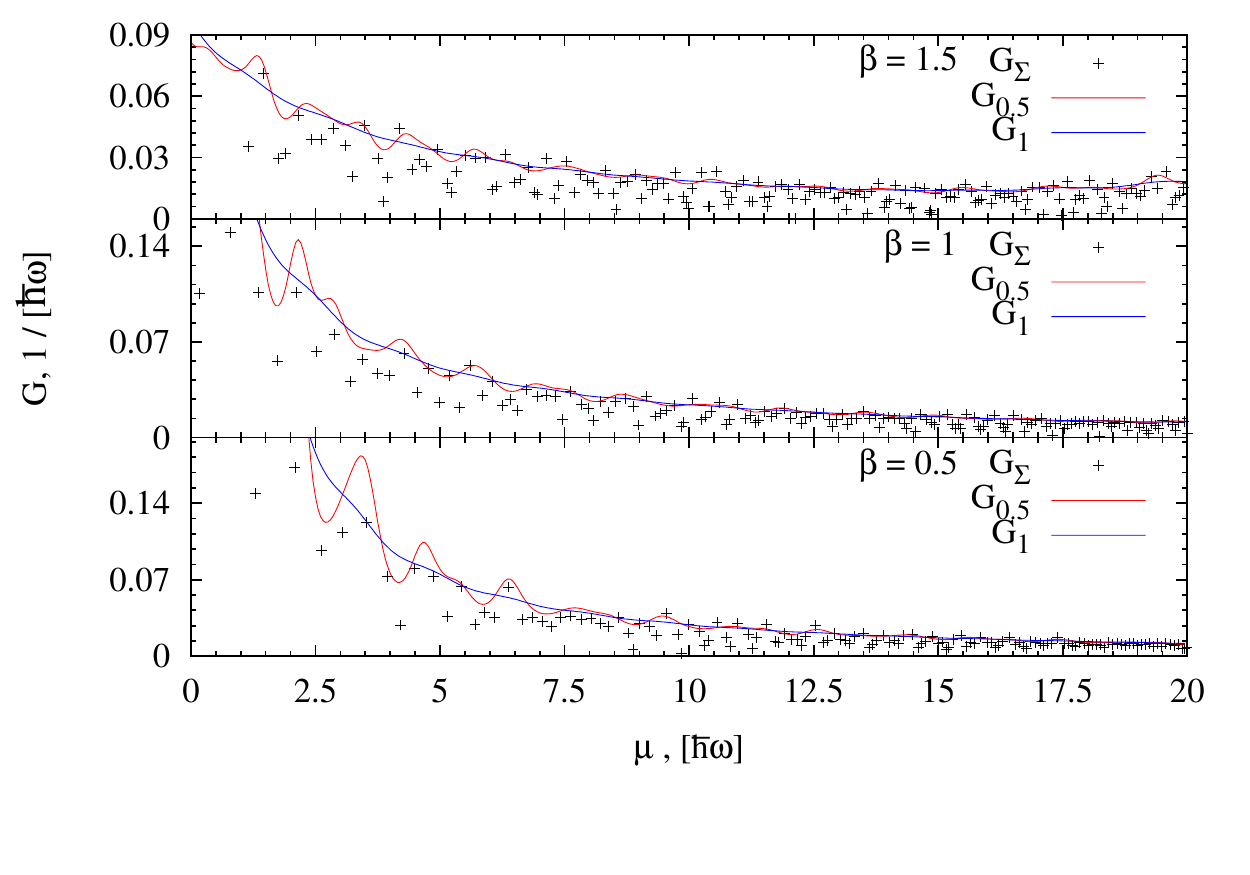}
\caption{The critical coupling constant, $G_c$ (in units of $\hbar\omega_y$), as a function of the chemical potential, $\mu$, for the spherical $\gamma=1$ case (upper panel) and deformed cases with the frequency ratios $\gamma = 2$ (middle panel) and $\gamma = 1.57$ (lower panel). The smearing parameter, $\delta$, and the dimensionless Rashba coupling parameter, $\beta$, are given in the panels. In the legende we 
indicate by $G_\Sigma$ results obtained by summation over the single-particle states, while $G_{0.5}$ and $G_1$ are obtained by 
smearing using the parameters $\delta=0.5$ and $\delta=1$ respectively. Here we set $\omega=\omega_y$.}
\label{figure6}
\end{figure}

In the top and middle panels of figure~\ref{figure6} we show the resulting $G_c$ obtained from Eqs.~\eqref{crit_coup}) and \eqref{crit_coup_int} for different $\beta$-values and deformations. The horizontal axis is the energy $\varepsilon$, and the discrete points are chosen as energies precisely between two single-particle levels. These points represent chemical potential values, $\mu$, and correspond to some uniquely given particle numbers (via the standard number equation). For degenerate levels this implies that at least one single-particle energy equals the value of the chemical potential, and consequently the critical strength must vanish according to Eq.~\eqref{crit_coup_int}.  The spherical oscillator with $\beta= 0$ is therefore not a suitable illustration, and we avoided this example in figure~\ref{figure6}.  The same problem arises for deformations where the frequency ratios have small integer values. The global behavior of $G_c$ is a decrease with $\mu$. This is most clearly seen by the average curves arising from Eq.~\eqref{crit_coup_int} with two different values of $\delta$. For $\delta = \hbar \omega$ the oscillations have essentially all disappeared, leaving only the decreasing average trend.  For a smaller $\delta$ value the still smooth oscillations tend to pick up and sometimes even reproduce the
microscopic behavior from Eq.~\eqref{crit_coup}. This is remarkable due to the simplicity in both derivation and resulting expression, Eq.~\eqref{crit_coup_int}. The dependence on the interaction interval $S$ is logarithmic and hence rather insignificant unless the absolute value is much smaller than the shell distance.

The microscopic behavior is seen in the discrete points where large variations appear corresponding to critical values from almost zero
and up to about $\hbar \omega_y/30$. The vanishingly small values correspond to black regions in the single-particle spectra of figure~\ref{figure1}, whereas the appreciable finite critical values correspond to the white regions. The behavior is qualitatively similar for different $\beta$-values and deformations. However, the particle numbers corresponding to small and large values of $G_c$ are completely different and strongly depending on the features of the underlying single-particle spectra. An approximate average relation between $\mu$ and $N$ can be seen in Eq.~\eqref{mu_and_N} with $\mu = \varepsilon_F$.

The curves from Eq.~\eqref{crit_coup_int} are overall at the upper limit of a genuine average. This is due to the systematical underestimate of the density of states at the chemical potential chosen to be exactly between two levels. The tendency of increasing the critical strength with $\beta$ is due to the overall increase of the density of states for any given particle number.  In general, the critical strengths are very fluctuating. This suggests that fine-tuning of particle number, Rashba coupling, and shape of the external field can lead to desirable but qualitatively different structures.

\section{Discussion}
We have shown that the single-particle spectrum for particles that 
are subject to a Rashba spin-orbit coupling in a deformed harmonic
trap is very dependent on the parameters and that spectral gaps
can be tuned by varying the Rashba coupling strength, a perpendicular 
Zeeman field and by changing the deformation of the trapping 
potential in the plane where the Rashba coupling acts. In particular, 
we find that the effect of deformation can be very different depending
on whether the ratio of the trapping frequencies in the plane is 
equal to ratios of small integers or if it is closer to an irrational 
number. This can change the spectrum from having an intricate 
structure with several characteristic gap sizes into a more evenly
spaced distribution that approximates a continuum.

In this study we have not taken interactions into account. However, it 
is certainly possible to infer some potential effects of our results on 
a system with more particles and with interactions. The simple case of 
a non-interacting Fermi gas becomes very intriguing with the spectra 
presented here. If the Fermi energy is located inside one of the pockets with no
states the system 
would be more stable than the case where
many single-particle states are close to the Fermi energy.
With the possibility of tuning this by external means we 
can explore this physics in a very direct manner. If we consider
a weak interaction in a mean-field picture like the BCS approach, 
the Rashba coupled case is extremely interesting as has 
been discussed recently \cite{vya2011,yu2011,gong2011,iskin2011,han2012,vya2012}.
From BCS theory in finite systems and/or in a trap we know that one needs a critical 
coupling strength to obtain a superfluid state (in free space the 
critical value goes to zero). This critical value is crudely speaking
inverse proportional to the density of states. In the present study we
have shown that it can vary strongly with the spin-orbit interaction 
strength and the deformation of the trapping potential.
The tunability demonstrated by our results imply 
that one can probe exotic superfluid states by using the deformation. 

For neutral single-species ultracold atoms the interactions are of 
van der Waals type and for the low experimental densities this can be 
considered of essentially zero-range. For a system of fermions with 
two internal hyperfine states this 
implies that one only has interaction in the spin singlet channel. The 
results presented here can be generalized to the two-particle case and 
one may then ask for the probability to find two fermions in the 
singlet state as the Rashba coupling strength, the Zeeman field and 
the deformation is varied. This will then tell us how the interactions
are suppressed by the trapping potential in the presence of a spin-orbit
coupling. In fact this is the first step toward the study of few-body 
physics in trapped system with non-abelian gauge potentials, and this 
is a direction that we will explore in the near future. Some open 
questions are the spectrum of two and three fermions in a trap with 
short-range interactions and its dependence on the deformation of 
the external confinement. Recent work has explored the case of 
few-boson problems in gauge fields in order to study strongly
correlated dynamics akin to quantum Hall systems \cite{julia2011,julia2012}.
We expect that fermions in different trapping geometries should also 
give rise to interesting strongly correlated states when subjected 
to external gauge potentials.

\paragraph*{Acknowledgements} This work was supported by the Danish 
Agency for Science, Technology, and Innovation under the 
Danish Council for Independent Research - Natural Sciences.

\end{document}